\documentclass[preprintnumbers]{revtex4}
\usepackage{amsmath}
\usepackage{amssymb}
\usepackage{graphicx}

\input{tcilatex}
\begin{document}

\date{\today }
\title{A short summary on the search of trineutron and tetraneutron}
\author{Roman Ya. Kezerashvili$^{1,2}$}
\affiliation{\mbox{$^{1}$Physics Department, New York
City College of Technology, The City University of New York,} \\
Brooklyn, NY 11201, USA \\
\mbox{$^{2}$The Graduate School and University Center, The
City University of New York,} \\
New York, NY 10016, USA }

\begin{abstract}
In light of a new experiment which claims an identification of tetraneutron 
\cite{Kisamori2016}, we discuss the results of experimental search of
trineutron and tetraneutron in different nuclear reactions. A summary of
theoretical studies for trineutron and tetraneutron within variety of
approaches such as variational methods, the method of Faddeev and
Faddeev-Yakubovsky equations, and the method of hyperspherical harmonics are
presented.
\end{abstract}

\maketitle

\bigskip The cataclysmic events that occur near the end of the life of a
star lead to one of only three possible final states: a white dwarf, a
neutron star, or a black hole. The mass of the star, particularly that of
the core, appears to be the primary factor in determining the final state. A
more massive star would need to be hotter to balance its stronger
gravitational attraction. While a star is burning, the heat in the star
pushes out and balances the force of gravity. When the star's fuel is spent,
and it stops burning, there is no heat left to counteract the force of
gravity. How much mass the star had when it died determines what it becomes.
Detailed calculations have shown that for star with mass less than about 1.4
times the mass of our sun electron degeneracy pressure permanently halts
collapse. White dwarfs are stable cold stars that are supported by electron
degeneracy pressure. Calculations show that stars that have between 1.4 and
3 times the mass of the sun implode into neutron stars that are the end
product of stellar evolution, and their outer core is composed of neutrons
at truly enormous densities. The central region of the neutron star is
supported by the degeneracy pressure of neutrons. A star with mass greater
than 3 times than of the sun gets crushed into a single point - a black hole.

At high density, when the sum of masses of a proton and electron, and Fermi
energy over $c^{2}$ exceeds the neutron mass, it is energetically favorable
to combine a proton and an electron into a neutron: $p+e^{-}\rightarrow
n+\nu _{e}.$ Both neutron and neutrino rich matter are produced at the core.
Since the mean free path of neutrinos is much smaller than the radius of
neutron stars neutrinos do not accumulate inside neutron stars. Therefore,
at higher densities, matter becomes more and more neutron-rich. A
progressive neutronization of matter at higher and higher densities makes a
lower energy state. An attractive pairing interaction between neutrons, can
couple them to form a state with integer spin and, therefore, paired
neutrons act like bosons. These "bosons" can form a condensate-like state in
which all of the bosons occupy the same quantum state and form a superfluid.
Just as the pairing of protons that are charged fermions forms a
superconductor. In that same general sense, we also can have
superconductivity and superfluidity in neutron stars. Thus, we can have
superconductivity and superfluidity in the outer core of neutron stars.
Superconductivity and superfluidity, if observed in neutron stars, could
tell us a lot about the pairing and hence inform us about aspects of nuclear
physics that are mighty difficult to get from laboratories. Information on
multineutron forces obtained in studies of multineutron systems is a
critical input into theories of neutron stars \cite{Demorest2010, Hiyama}.
Therefore, the study of dineutrons, trineutrons, tetraneutrons as well as
multineutrons and neutron drops is important for understanding the structure
and processes in neutron stars and any experimental evidence for the
existence of such states has implications for research into neutron stars.

Below, I give a short summary and present the status of studies of
trineutrons and tetraneutrons.

A simple fact has now been established: all nuclei that are heavier than the
hydrogen nucleus are made up of both protons and neutrons. The question then
arises as to whether a nuclei made up of only neutrons or protons can exist.
On the basis of current knowledge the theoretical answer is probably ...
Well, let's discuss this!

More than half a century experimental search and theoretical investigation
has continued to focus on atomic nuclei consisting only of neutrons. A
recently reported observation of the tetraneutrons \cite{Kisamori2016} and
some theoretical results, however, revives old questions: do dineutrons,
trineutrons and multineutrons nuclei exist? Can a nucleus be made up of
neutrons only? Does neutron matter exist? The existence of a bound
dineutron, multineutrons nuclei, neutron drops and neutron matter is of
great importance, as it would challenge our understanding of nuclear
few-body systems and the evolution of the universe. The answers to these
questions would most certainly require a revision of modern realistic models
of the nucleon-nucleon force and three-nucleon interaction and more over
introduce a four-nucleon interaction.

A free neutron decays into a proton, an electron, and an antineutrino, which
is associated with an electron. The time of this decay is about 1000
seconds. In other words, a free neutron may exist for only about 16 minutes.
What about the existence of the system of two bound neutrons known as a
dineutron? Searching resonances and bound states in a system of two neutrons
is a well defined problem and numerically well under control. The two
neutron resonances are associated with the poles of the $S$-matrix, which
are embedded in the fourth-quadrant of the complex $k$ plane. They are
solutions of the time-dependent Schr\"{o}dinger equation without the
incoming wave and the outgoing wave increasing exponentially at infinity. In
Ref. \cite{Pieper2003} it was mentioned that nonrealistic Volkov potentials 
\cite{Volkov1965} do have bound dineutrons. However, these potentials are
not realistic; they produce bound $^{2}$n, with the same binding energies as
their deuterons; they have no tensor or $LS$ terms; and they cannot
reproduce modern phase shift analyses in any partial wave \cite{Pieper2003}.
Theoretical calculations show no existence of the resonance or bound states
in the system of two neutrons for all the existing realistic and
phenomenological models of the nucleon-nucleon interaction. Experimental
searches of the dineutron have been also performed using different nuclear
reactions but no evidence for the existence of the $^{2}$n has been found.
Thus, today we are confident that dineutrons do not exist but could very
nearly exist: a slight increase in the attraction between the two particles
would result in a bound structure, the dineutron being formed. However, in
neutron-rich matter like a neutron star where the density 3-times as much as
the normal nuclear density would nucleon-nucleon interaction modified so
that brings two neutrons to be bound? This question needs to be addressed.
Existence of dineutrons in neutron stars could bring a new mechanism of
superfluidity of neutron matter like an excitonic one in condensed matter
physics (see, for example, Ref. \cite{BermanKezPRB2012Polariton}). Research
into the possibility that nuclei have more than two neutrons shows that,
very often, adding a further neutron increases the stability of the
structure. The question then arises as to whether a neutron system made up
of more than two neutrons could exist.

A three-neutron resonance has not yet been firmly established. The weight of
early experimental evidence reviewed in Ref. \cite{Fiarman1975} is strongly
against the existence of a bound state of the three-neutron system, and only
controversial evidence of a three-neutron resonance was cited. The situation
up to 1987 has been reviewed in the compilation \cite{Tilley1987}. Reference 
\cite{Ajdacic1965} reported the possible existence of the trineutron through
the reaction $^{3}$H$(n,p)^{3}$n. Later, the same reaction was studied in
Refs. \cite{Thotnton1966} and \cite{Cernigoi81PiP} (see references herein)
no evidence for the existence of the trineutron was found. Searches for a
bound state of the three neutron were conducted in reactions: $^{3}$H$(\pi
^{-},\gamma )^{3}$n, $^{3}$He$(\pi ^{-},\pi ^{+})^{3}$n, $^{4}$He$(\pi
^{-},p)^{3}$n, and $^{7}$He$(\pi ^{-},^{4}$He$)^{3}$n induced by negative
pions, and ion collision reactions such as $^{7}$Li($^{7}$Li, $^{11}$C)$^{3}$%
n, $^{7}$Li($^{11}$B, $^{15}$O)$^{3}$n and $^{2}$H($^{12}$C, $^{13}$N)$^{3}$%
n.

Based on the method of description of true $3\rightarrow 3$ elastic
scattering within the method of hyperspherical function \cite%
{Kezerashvili1983, Kezerashvili2001} were studied the formation of three
neutrons in the processes $^{3}$He$(\pi ^{-},\pi ^{+})^{3}$n \cite%
{Kezerashvili84, Kezerashvili85}, charged-pion photo-production on $^{3}$H, 
\cite{Kezerashvili1984PhotoProdPi} and $\mu ^{-}-$capture by triton \cite%
{Kezerashvili1984muon, Kezerashvili1993Pisa}. As of yet, none of these
reactions have provided evidence for a bound trineutron. The most intensive
search for the prediction of a bound trineutron has been performed using a
pionic double charge exchange reaction $^{3}$He$(\pi ^{-},\pi ^{+})^{3}$n.
An investigation \cite{Grater1999} of the process $^{3}$He$(\pi ^{-},\pi
^{+})^{3}$n found no evidence of the existence of the $^{3}$n or resonance
state of three neutrons. Earlier study \cite{Sperrinde1970} pointed to
resonance in the three neutrons. However, the resonance behavior can be
explained by the final state interaction of three neutrons in continuum
spectrum as was demonstrated in Refs. \cite{Kezerashvili84, Kezerashvili85}.
The double charge exchange process on $^{3}$He was also investigated in Ref. 
\cite{Stez1986}, which while criticizing previous work \cite{Sperrinde1970}
pointed again to a three-neutron resonance around 12 MeV excitation. For a
trineutron the bound state has been studied extensively in the last four
decades resulting in a numerically precise solution of the Faddeev equations
in momentum and coordinate space, and using the hyperspherical functions
method. Some of the contradictory aspects of Faddeev calculations which
predicted a trineutron and variational calculations which did not, was
addressed in Ref. \cite{Glockle1979} using the Faddeev method with the Reid
potential and pointed out that interaction potential must be enhanced by
about a factor 4 to bind $^{3}$n near zero energy. The possibility of a
trineutron bound state is investigated in \cite{Sunami1980} by solving the
Faddeev equation in the coordinate space and have considered possible states
of the trineutron including even the $^{3}P_{2}-^{3}F_{2}$ force. The author
concluded that all states of $^{3}$n are unbound for the Reid soft core
potential. \ Independent of the theoretical framework, such as the Faddeev
formalism \cite{Witala1999, Glockle2002}, the method of the hyperspherical
functions or variational calculations, most theoretical works do not predict
a bound $^{3}$n state in the three-neutron system. However, it has been
stressed in Ref. \cite{Baz1972} that subtle changes in the nucleon-nucleon
potential, which would not affect results from phase shift analyses, may
lead to bound neutronic nuclei. To summarize, although the double charge
exchange reaction of negative pions on $^{3}$He nucleus has been examined at
various incident energies of pion, from the analysis of the invariant mass
spectra for three neutrons, no evidence for the bound trineutron has been
found. However, a calculation \cite{Csoto1996} predicts a resonance state
with the width of 13 MeV in the three-neutron system. Although such a
resonance would easily fit early interpretation of data on pionic double
charge exchange on $^{3}$He \cite{Sperrinde1970} (this observation was
supported by measurements reported in Ref. \cite{Stez1986} in 1986), more
recent investigations \cite{Grater1999, GraterPL1999, Yuly1997} of this
process do not give any experimental evidence for it. All reactions
mentioned above are plagued by the fact that although there is an
expectation that a $^{3}$n system is produced, at least one additional
strongly interacting particle is in the final state. The reactions $\pi
^{-}+^{3}$H$\rightarrow $ $^{3}$n+$\gamma $ and $\mu ^{-}+^{3}$H$\rightarrow 
$ $^{3}$n+$\nu _{\mu }$ have only an extra photon or neutrino in the final
state. Additional interest in these reactions is in the $^{3}$H-$\pi $ and $%
^{3}$H-$\mu $ atomic physics since the tritium is unique because the pion
and muon is exclusively absorbed from 1$s$ orbit. Theoretically the $s$%
-state radiative $\pi $-capture or $\mu $-capture transition rates can be
calculated quite accurately. Refs. \cite{3HpiGamma1976, 3HpiGamma1980}
reported trineutron mass spectrum that was observed in the stopped pion
radiative capture reaction $^{3}$H($\pi ^{-}$,$\gamma $)3n by measuring the
photon energy with a high-resolution spectrometer. However, as also in
earlier studies, no indication of a bound trineutron is observed. The
noticeable difference in $\mu ^{-}-$capture by a tritium nucleus obtained in
calculations \cite{Kezerashvili1984muon, Kezerashvili1993Pisa} by taking
into account the final state interaction between three neutrons and a
plane-wave approximation increases the interest in the experimental study of
the channel of the complete break up of $^{3}$H in $\mu ^{-}-$capture.

Several more recent experiments have strengthened the evidence against the
bound trineutron and have failed to discover a resonance structure that
cannot be otherwise explained. The study \cite{Lazauskas2005} shows that
realistic nucleon-nucleon interaction models exclude any possible
experimental signature of three-neutron resonances. Thus, today there is no
unambiguous answer for the existence of the three-neutron nucleus. Apart
from these aspects the question of whether multineutron systems exist is of
principal interest by itself.

The search for the tetraneutron has been going on for more than fifty years 
\cite{Zelevinsky2016}. In a compilation of information on A = 4 nuclei given
in \cite{TilleyA4} no firm evidence for bound states of $^{4}$n is
presented. This year, a candidate resonant tetraneutron state with the
energy of 0.83$\pm $ 0.65(stat)$\pm $1.25(syst) MeV above the threshold of
four-neutron decay has been found in the missing-mass spectrum obtained in
the double charge exchange reaction $^{4}$He($^{8}$He,$^{8}$Be) at 186 MeV/u 
\cite{Kisamori2016}. The experiment was performed at the RI Beam Factory at
RIKEN. Previously, in experiments, the system of four bound neutrons $^{4}$n
was searched through using heavy-ion transfer reactions such as $^{7}$Li($%
^{11}$B, $^{14}$O)$^{4}$n \cite{Belozyorov1988}, $^{7}$Li($^{7}$Li, $^{10}$C)%
$^{4}$n \cite{Aleksandrov1988}, and the pion double charge exchange reaction 
$^{4}$He$(\pi ^{-},\pi ^{+})^{4}$n. Early measurements of the $^{4}$He$(\pi
^{-},\pi ^{+})^{4}$n reaction carried out in search of evidence for $^{4}$n
are summarized in the compilation \cite{Fiarman19754n}. No bound $^{4}$n was
detected in these early works. Later the momentum spectrum from the pion
double charge exchange reaction was measured in Ref. \cite{Urgan1984} in a
search for $^{4}$n. Note, however, that the theoretical study \cite%
{Kezerashvili1986} reported that the final-state interaction in the
four-neutron system in continuum spectrum is so strong that the tetraneutron
could not be observed in the kinematic region explored in Ref. \cite%
{Urgan1984}. Pion spectra and total cross sections for pion double charge
exchange were also measured in Refs. \cite{Stetz1981, Kinney1986,
Gorringe1989} for different incident pion energies. No evidence for $^{4}$n
was obtained. Several attempts have been made to find a bound tetraneutron
system by using a uranium fission reaction \cite{Schiffer1963,
Cierjacks1965, Kiev} and the experimental observation of $^{4}$n was claimed
in the interaction of 100 MeV $\alpha -$particle with uranium nucleus in
Ref. \cite{Kiev}.

Several theoretical studies of pion double charge exchange on $^{4}$He have
been reported. In Ref. \cite{Gibbs1977} cross sections were calculated in a
model in which two single charge exchange scatterings occur. The reaction $%
^{4}$He$(\pi ^{-},\pi ^{+})^{4}$n was studied in the framework of a
four-body hyperspherical basis method in Ref. \cite{Kezerashvili1980,
Kezerashvili1981, Kezerashvili85} but existing experimental data were
interpreted without bound or resonance state of four neutrons. The review of
the studies of the four neutrons produced by $\pi ^{-}$ double charge
exchange reaction on $^{4}$He nucleus is given in Ref. \cite%
{KezerashviliEChaYa1985}. No bound tetraneutron was found in Ref. \cite%
{Gorbatov1989} within the angular potential functions method, in Ref. \cite%
{Vagra1995} using the stochastic variational method and in Refs. \cite%
{Badalyan1985} and \cite{Sofianos1997} within the hyperspherical functions
approach. In \cite{Sofianos1997} using a local $S$-wave $nn$-potential
possible bound and resonant states of three and four neutrons systems are
sought as zeros of three- and four-body Jost functions in the complex
momentum plane. It is found that zeros closest to the origin correspond to
subthreshold ($3n$)$\frac{1}{2}^{-}$ and ($n$)$0^{+}$ resonant states. In
contrast, calculations within the hyperspherical functions method led the
authors in Ref. \cite{Gutich1989} to the conclusion that the tetraneutron
may exist as a resonance only for the $NN$ potential that binds the
dineutron.

In the new millennium, an experimental search and theoretical study of
tetraneutron transitioned to a new phase. In 2002 an international team led
by physicists from the Particle Physics Laboratory of Caen, have presented
in Ref. \cite{Marques2002} experimental results suggesting the existence of
a bound tetraneutron. These results have been obtained by using the exotic
beams of the French national large heavy-ion accelerator in Caen and by
studying the breakup reaction of $^{14}$Be into $^{10}$Be and bound
tetraneutron. The heavy-ion transfer reactions, and the pion double charge
exchange reaction require considerable reconfiguration of the target nuclei
and should be strongly suppressed. In contrast, the nucleus $^{14}$Be
consists of a strongly bound core $^{10}$Be and four weakly bound neutrons
that could form a tetraneutron-like configuration, which might be shaken off
in the $^{14}$Be breakup. In experiment \cite{Marques2002} events were
observed that exhibit the tetraneutron cluster liberated in the breakup of $%
^{14}$Be. In particular, the fragmentation channel $^{10}$Be$+^{4}$n was
observed and the $^{4}$n was described as a bound four neutrons system. The
lifetime suggested by this measurement would indicate that the tetraneutron
is a stable particle. In 2016, the resonant tetraneutron state was found in
the reaction $^{4}$He($^{8}$He,$^{8}$Be) \cite{Kisamori2016}. If confirmed,
these discoveries, which would call into question current theoretical
models, will have major repercussions in the field of nuclear physics.
Therefore, the confirmation of such an exotic state with a different
reaction is important. Motivated by the recent observation of tetraneutron
candidates at RIKEN \cite{Kisamori2016}, there is a proposal to investigate
high energy forward scattering pion double-charge-exchange reaction on $^{4}$%
He at J-PARC \cite{Fujioka2016}. As opposed to the old experiments at LAMPF
and TRIUMF, a rather high-energy pion beam, around 850 MeV, will be utilized
with the benefit of reduction of the irrelevant background.

The existence of the bound tetraneutron system was also discussed in
theoretical studies \cite{Timofeyuk, Zelevinsky2003, Zhukov2004,
Lazauskas20054n, Grinyuk2006, Lashko2008}. In Ref. \cite{Zelevinsky2003} it
was proposed that, if tetraneutron existed, it could be formed by a bound
state of two dineutron molecules. The possibility for a tetraneutron to
exist as a low-energy resonance state was studied in Ref. \cite{Lashko2008}.
In \cite{Timofeyuk} the hyperspherical functions method and realistic
nucleon-nucleon interactions have been used to argue against the existence
of a tetraneutron. It was pointed out that due to the small probability for
a pair of neutrons to be in the singlet even state, the two-body nuclear
force cannot by itself bind four neutrons, even if it could bind a
dineutron. In Ref. \cite{Lazauskas20054n} Faddeev-Yakubovsky equations have
been solved in configuration space using a realistic Reid 93 $nn$
interaction to explore the possible existence of four-neutron resonances
close to the physical energy region. The authors used two methods, namely,
complex scaling and analytical continuation in the coupling constant to
follow the resonance pole trajectories, which emerge out of artificially
bound tetraneutron states. Four-neutron states evolving from the forced
bound state end up in the unphysical third energy quadrant with negative
real energy parts, thus having little effect on any physical process. It was
demonstrated that the tetraneutron physics is entirely determined by $nn$ $S$
waves, namely $^{1}S_{0}$ one, which is controlled by the experimentally
measurable $nn-$scattering length \cite{Lazauskas20054n}. 

An unrealistic modification of the nucleon-nucleon force, three-nucleon
force or introduction of unrealistic four-nucleon forces would be needed to
bind a tetraneutron. It is also important to mention that Ref. \cite%
{Pieper2003} shows that it does not seem possible to change any modern two-
and three-nucleon interaction to bind a tetraneutron without destroying many
other successful predictions of these interactions. This means that, should
recent experimental claims \cite{Marques2002, Kisamori2016} of a
tetraneutron be confirmed, our understanding of nuclear forces will have to
be significantly changed. Recently, in Ref. \cite{Hiyama2016} a possibility
of generating a four-neutron resonance with introducing of a
phenomenological $T=3/2$ three-neutron force in addition to a realistic $NN$
interaction (Argonne AV8$^{^{\prime }}$ version of the $NN$ potential) is
considered. Using these interactions the \textit{ab initio }solution of the 4%
$n$ Schr\H{o}dinger equation is obtained using the complex scaling method
with boundary conditions appropriate to the four-body resonances by
inquiring what the strength should be of the three-neutron force to generate
such a resonance. Considering the most natural way to enhance a tetraneutron
system near the threshold through an additional attractive isospin $T=3/2$
term in the three-body force, but this modification was found to be
inconsistent with other well-established nuclear properties and low energy
scattering data \cite{Hiyama2016}. Therefore, it was demonstrated that a
four-neutron resonance requires unrealistic changes in the three-neutron
force to have observable resonance features. However, recent theoretical
calculation \cite{Shirokov2016} within various \textit{ab initio} approaches
with nucleon-nucleon interaction JISP16 \cite{ShirokovJISP16} predicts a low
energy $^{4}$n resonance near 0.8 MeV above threshold with a width of about
1.4 MeV. The calculation is performed without altering any of the properties
of the $NN$ interaction. The reported result is compatible with the recent
experimental finding \cite{Kisamori2016} for a resonant structure.

The negative result of numerous searches for trineutron and tetraneutron
does not exclude the existence of heavier neutron clusters. As for
experimental searches of the other light bound multineutron systems, 
light nuclear stable multineutrons among products of the fission of $^{238}$%
U nuclei that is induced by 62 MeV alpha particles have been found by the
activation method with the $^{88}$Sr \cite{Novatskii2012} and $^{27}$Al \cite%
{Novatskii2013} isotopes, respectively. These multineutrons have been
detected by characteristic $\gamma $ rays that correspond to the formation
of a beta active $^{92}$Sr nucleus and the formation of the latter one was
associated with at least a four-neutron-transfer reaction involving a
nuclear stable multineutron: $^{88}$Sr($^{x}$n, $(x-4)$n)$^{92}$Sr \cite%
{Novatskii2012}. In case of $^{27}$Al the multineutrons have been detected
by characteristic $\gamma $ rays emitted by the nuclei from the beta decay
chain $^{28}$Mg$\rightarrow ^{28}$Al$\rightarrow ^{28}$Si. The $^{28}$Mg
parent nucleus could be formed in the $^{27}$Al$+^{x}$n$\rightarrow ^{28}$Mg$%
+$($x$ -- 2)np process. The results of these two independent experiments 
\cite{Novatskii2012, Novatskii2013} indicate that nuclear stable
multineutrons (most likely, $^{6}$n) are emitted from the alpha particle
induced ternary fission of $^{238}$U.

The most recent study \cite{Bystritskiy} with the duration of the experiment
of 112.66 days indicates that in the cluster decay of $^{238}$U there is a $%
^{6}$n and (or) $^{8}$n via reactions: $^{238}$U$\rightarrow $ $^{6}$n$+$ $%
^{232}$U and $^{238}$U$\rightarrow $ $^{8}$n$+$ $^{230}$U. Nuclei $^{232}$U
and $^{230}$U as a result of a chain of alpha, beta decays turn into $^{208}$%
Pb and $^{206}$Pb nucleus, respectively. The probability of $^{6}$n and (or) 
$^{8}$n emission of nuclei $^{238}$U with respect to the probability of $%
\alpha -$decay mode are determined and the upper limits on the 90\%
confidence level are the following: $\lambda $($^{6}$n)$/\lambda _{\alpha
}\leq $ $9.3\times 10^{-9}$ and (or) $\lambda $($^{6}$n)$/\lambda _{\alpha
}\leq $ $3.6\times 10^{-10}$, where $\lambda $ is the corresponding decay
constant.

On the other side, theoretical investigations of multineutrons have been
carried out for a system of six, eight and ten bound neutrons using the
hyperspherical functions method with different nucleon-nucleon interactions 
\cite{Timofeyuk}. Results of calculations show no bound states for these
neutron systems. Theorists are also studying neutron drops \cite{Pieper96,
Pieper2013}. Neutron drops are collections of neutrons held together by both
an external nuclear well and the interaction between neutrons. The
properties of these drops can be used as "data" for fitting simpler
effective interaction models that are employed in the study of large
neutron-rich nuclei, the crusts of neutron stars, and neutron matter.

Today there is no unambiguous answer for the existence of the trineutron as
a bound or resonance state. There are three claims of experimental
observation of $^{4}$n: one in a fission reaction \cite{Kiev}, and the
recent two in the breakup reaction of $^{14}$Be into $^{10}$Be and $^{4}$n 
\cite{Marques2002}, and in the double-charge-exchange reaction $^{4}$He($%
^{8} $He,$^{8}$Be) \cite{Kisamori2016}. However, in theoretical studies,
except very recent one \cite{Shirokov2016} with the JISP16 potential \cite%
{ShirokovJISP16}, no evidence for $^{4}$n was obtained within the existing
modern two- and three-nucleon interaction. If the experimental discoveries 
\cite{Marques2002} and \cite{Kisamori2016} as well as the obtained evidences 
\cite{Novatskii2012, Novatskii2013, Bystritskiy} in support of the existence
of neutron nuclei whose number is not lower than six would be confirmed,
this would call into question current theoretical models of nuclear forces.
Interestingly enough, such consideration has already started \cite%
{Hiyama2016}.

\acknowledgments The author thanks R. Lazauskas, J. Carbonell, N. K.
Timofeyuk, H. Fujioka and S. Shimoura for valuable discussions.

\end{document}